\documentstyle[draft,psfig]{mn} 

\newif\ifAMStwofonts

\ifoldfss
  \ifCUPmtlplainloaded \else
    \NewTextAlphabet{textbfit} {cmbxti10} {}
    \NewTextAlphabet{textbfss} {cmssbx10} {}
    \NewMathAlphabet{mathbfit} {cmbxti10} {} 
    \NewMathAlphabet{mathbfss} {cmssbx10} {} 
  \fi
  \ifAMStwofonts
    \ifCUPmtlplainloaded \else
      \NewSymbolFont{upmath} {eurm10}
      \NewSymbolFont{AMSa} {msam10}
      \NewMathSymbol{\upi}     {0}{upmath}{19}
      \NewMathSymbol{\umu}     {0}{upmath}{16}
      \NewMathSymbol{\upartial}{0}{upmath}{40}
      \NewMathSymbol{\leqslant}{3}{AMSa}{36}
      \NewMathSymbol{\geqslant}{3}{AMSa}{3E}

      \let\geq=\geqslant 
    \fi
  \fi
\fi 

\ifnfssone
  \newmathalphabet{\mathit}
  \addtoversion{normal}{\mathit}{cmr}{m}{it}
  \addtoversion{bold}{\mathit}{cmr}{bx}{it}
  \newmathalphabet{\mathbfit} 
  \addtoversion{normal}{\mathbfit}{cmr}{bx}{it}
  \addtoversion{bold}{\mathbfit}{cmr}{bx}{it}
  \newmathalphabet{\mathbfss} 
  \addtoversion{normal}{\mathbfss}{cmss}{bx}{n}
  \addtoversion{bold}{\mathbfss}{cmss}{bx}{n}
  \ifAMStwofonts
    \ifCUPmtlplainloaded \else
      \UseAMStwoboldmath
      \makeatletter
      \new@mathgroup\upmath@group
      \define@mathgroup\mv@normal\upmath@group{eur}{m}{n}
      \define@mathgroup\mv@bold\upmath@group{eur}{b}{n}
      \edef\UPM{\hexnumber\upmath@group}
      \new@mathgroup\amsa@group
      \define@mathgroup\mv@normal\amsa@group{msa}{m}{n}
      \define@mathgroup\mv@bold\amsa@group{msa}{m}{n}
      \edef\AMSa{\hexnumber\amsa@group}
      \makeatother
      \mathchardef\upi="0\UPM19
      \mathchardef\umu="0\UPM16
      \mathchardef\upartial="0\UPM40
      \mathchardef\leqslant="3\AMSa36
      \mathchardef\geqslant="3\AMSa3E

      \let\geq=\geqslant 
    \fi
  \fi
\fi 

\ifnfsstwo
  \DeclareMathAlphabet{\mathbfit}{OT1}{cmr}{bx}{it}
  \SetMathAlphabet\mathbfit{bold}{OT1}{cmr}{bx}{it}
  \DeclareMathAlphabet{\mathbfss}{OT1}{cmss}{bx}{n}
  \SetMathAlphabet\mathbfss{bold}{OT1}{cmss}{bx}{n}
  \ifAMStwofonts
    \ifCUPmtlplainloaded \else
      \DeclareSymbolFont{UPM}{U}{eur}{m}{n}
      \SetSymbolFont{UPM}{bold}{U}{eur}{b}{n}
      \DeclareSymbolFont{AMSa}{U}{msa}{m}{n}
      \DeclareMathSymbol{\upi}{0}{UPM}{"19}
      \DeclareMathSymbol{\umu}{0}{UPM}{"16}
      \DeclareMathSymbol{\upartial}{0}{UPM}{"40}
      \DeclareMathSymbol{\leqslant}{3}{AMSa}{"36}
      \DeclareMathSymbol{\geqslant}{3}{AMSa}{"3E}

      \let\geq=\geqslant 
    \fi
  \fi
\fi 

\ifCUPmtlplainloaded \else
  \ifAMStwofonts \else 
    \def\upi{\pi}
    \def\umu{\mu}
    \def\upartial{\partial}
  \fi
\fi

\title[Dynamical Merger Trees]{Dynamical consequences of CDM merger trees}

\author[X. Hernandez \& W.~H. Lee]
{Xavier Hernandez and William H. Lee \\
Instituto de Astronom\'{\i}a, Universidad Nacional Aut\'{o}noma
de M\'{e}xico, Apdo. Postal 70--264, Cd. Universitaria, 04510
M\'{e}xico D.F.\\
}

\pagerange{\pageref{firstpage}--\pageref{lastpage}}
\pubyear{2002}

\begin{document}

\maketitle

\label{firstpage}


\begin{abstract}
Within the context of the standard structure formation scenario, massive
present day
elliptical galaxies are sometimes thought of as the result of a major merger
of spiral systems. Through extensive SPH simulations of merging
spirals, we have explored these processes with the aim of quantifying
their relaxation times. This is important, as it sets a minimum time
interval between the onset of a merger, and the appearance of an
elliptical galaxy. We then compare this constraint with predictions of
the hierarchical scenario, computed through Press-Schechter merger
trees. We find evidence for elliptical systems which appear not
to have been formed by a major merger of spirals.

\end{abstract}
   
\begin{keywords}
galaxies: formation - galaxies: evolution - galaxies: elliptical and
lenticular, cD - galaxies: interactions - cosmology: theory.
\end{keywords}


\section{Introduction} \label{intro}

In cosmology, the inflationary paradigm has not only provided an
elegant answer to several disturbing problems of the standard Big Bang
such as the horizon and the ``flatness'' problems, but also provides a
mechanism for generating the primordial fluctuations from which we
expect astrophysical structures to be formed.  In most simple
inflationary models, the initial spectrum of these fluctuations is a
power law of the scale, and there is total absence of phase
correlation among different scales.  These properties lead to the
hierarchical scenario of structure formation, where small scale
objects merge continuously to produce increasingly larger ones as
time goes by (e.g. White \& Rees 1978; White et al. 1987).

The predictions of this model at large scale have been highly
successful in matching the observed universe. N-body simulations of
galactic clusters and super clusters match equivalent observed systems
remarkably well. However, at galactic and sub-galactic scales
considerable debate remains. Is the centrally concentrated dark matter
density profile obtained from simulations of galactic dark haloes
representative of the constant density cores seen in real galaxies or
not (e.g. de Blok \& McGaugh 1997; Firmani et al. 2001; Gnedin \& Zhao
2002)? Is the level of substructure predicted by simulations at
galactic levels compatible with the abundance of satellite systems in
large galaxies, or are too many satellites being predicted (e.g. Moore
et al. 1999; Ghigna et al. 2000)? Are the sizes of large disk galaxies
compatible with the loss of angular momentum which models predict,
mostly as a consequence of the extensive merging regime these systems
should have been subject to, which should probably have heated the
disks beyond observed constraints (e.g. Navarro \& Steinmetz 2000)?

It is clear that the assumptions going into the hierarchical model, at
galactic and sub-galactic regions, remain subject to serious
doubt. This is not surprising, since they result from large
extrapolation of the direct analysis of primordial fluctuations
performed at a much larger scale range, mostly through the study of
the cosmic microwave background.

In this paper we shall examine another of the predictions of the
hierarchical clustering scenario at the galactic level, namely the
formation of massive elliptical galaxies through the merger of spiral
systems. Examples of this proposal can be found in: Kauffmann (1996),
Baugh et al. (1996), Somerville et al. (2001), Benson et al. (2002), Steinmetz \& Navarro (2002)
and Khochfar \& Burkert (2003). That spiral galaxies sometimes collide and merge is an observational
fact, that the remnant closely resembles an elliptical galaxy, has been proven repeatedly through
numerical simulations, nevertheless the above do not constitute proof of the main formation
mechanism for elliptical galaxies being through major mergers of spirals.

Through detailed Smooth Particle Hydrodynamics (SPH)
simulations of the merger of two spirals, we will estimate the
relaxation times for the process. Once this has been established,
including a study of the variations expected given the extensive
parameter space available to such a merger, and once the uncertainties
in the initial conditions are considered, a consistency check for the
hypothesis is available.

Simulations of galactic mergers have been carried out many times
before, but mostly aimed at obtaining very particular information, and
rarely situated in a cosmological context, as what we attempt
here. For example, Hibbard \& Mihos (1995) explore merger remnant
morphology and star formation, Barnes (2002) studies the formation of
gas disks within merger remnants, Bendo \& Barnes (2000) explore the
distribution of line of sight velocities in merger remnants and
remnant morphologies.  Burkert \& Naab (2003) study the process of elliptical galaxy
formation through the merger of disks, but without including the dissipative gas component,
an ingredient which we find here to be of relevance. We deemed it necessary to repeat the
experiment, paying particular attention to setting up the initial
conditions in a cosmologically justified manner, as well as exploring
the explicit dependence of the final relaxation times on the ample
configuration parameter space. For example, Ellis (2001) remarks on the necessity
of cosmologically motivated merger simulations, to turn close galaxy pair statistics
into merger rates.

Merger trees for large ellipticals can be constructed analytically,
through the extended Press-Schechter formalism, and hence we have a
prediction of the time elapsed between the last major merger and the
present, for a galaxy of a given mass. This can be repeated, and given
a redshift of observation, the hierarchical merger scenario predicts
(on average) how far back in time the last major merger took
place. This can be compared to the dynamical estimates of the
relaxation time for the mergers. If the latter proves larger then the
former, the theory needs revising.

In Section 2 we describe the numerical scheme used to model a
collision between spiral galaxies, as well as the different
configurations tested. Section 3 gives the results of the different
simulations, giving the relaxation time criterion we wanted to
establish.  In Section 4 we calculate the Press-Schechter merger
trees for elliptical galaxies, and compare to the relaxation time
criterion of the previous section. Finally, our conclusions are given
in Section 5.

\section{Merger simulations}

\subsection{Numerical modeling} \label{method}

As stated in the introduction, we seek to obtain an estimate of the
relaxation timescales for the merger of two spirals to result in a
relaxed elliptical galaxy. Wanting to make a stringent comparison with
models of structure evolution, we shall model the formation of a high
redshift elliptical observed at $z \sim 1.0$. As we shall see, this
implies starting the simulation at $z \sim 1.5$.

In attempting to model numerically the merger of two spiral galaxies,
one must pay close attention to the gaseous component of the disk. It
is this gas fraction through which most of the dissipation will take
place, the other two components, stars and dark matter, being
non-collisional. We hence used the method known as Smooth Particle
Hydrodynamics (SPH), developed by Lucy (1977) and Gingold \& Monaghan
(1977), see Monaghan (1992) for a review. SPH is a Lagrangian method
ideally suited for complicated three-dimensional flows, and has been
used in a variety of astrophysical applications. We have specifically
used the public code GADGET, developed by Springel et al. (2001) for
galactic and cosmological simulations. This code allows us to trace
shock fronts and other hydrodynamical features of the gas with great
accuracy, as well as including an efficient N-body routine, to follow
the dynamics of the non-collisional components. Further the code is
fully parallelized, allowing it to run on a large number of computers
simultaneously, as was done in our case.

In setting up the spiral galaxies which are to collide, the first
thing is to construct each galaxy in isolation. We
include three components, a dark matter halo, a stellar disk and a
gaseous disk, co-planar to the stellar one. In some variants, a
stellar bulge was also added.

For the dark matter halo of each galaxy we chose a King sphere (King
1966). This has the advantage of including a fully self consistent
distribution function, with the density profile being a solution to a
Boltzmann and a Poisson equation. In this way we set up a ``live''
halo, which will respond to the formation of the disk in its centre
and, as the galaxies approach, to tidal effects and all other dynamics
of the merger. Although cosmological simulations (e.g. Navarro, Frenk
\& White 1996, Ghigna et al. 1998) result in dark matter halos which are much more
centrally concentrated than King spheres, direct observations of Dwarf
Irregulars and low surface brightness (LSB) galaxies seem to imply the
existence of constant density cores in the centers of galactic dark
matter halos (e.g. Burkert \& Silk 1997; de Blok \& McGaugh 1997),
which are inconsistent with cosmological profiles, even taking into
account possible expansions of the core due to efficient ejection of
the central baryonic component (Gnedin \& Zhao 2002) or observational effects (de Blok et al. 2003).  
In the dwarf spheroidals of the Milky Way, through velocity dispersion
studies, Lokas (2002) finds that a core is statistically preferred to a cusp
for the dark matter distribution, the same that results from a detailed
galaxy velocity dispersion study in the Coma cluster (Lokas \& Mamon 2003).
The situation in clusters though, is not as clear as in galaxies, e.g. 
through strong lensing, Tyson et al. (1998) 
find that the dark matter halo in  CL 0024+1654 is well represented by a structure having a central core, 
while van der Marel et al. (2000) find that CNOC1 redshift data for several clusters imply
velocity dispersions and mass profiles consistent with NFW profiles.

Further, Hernandez \& Gilmore (1998) showed that King halos produce rotation
curves which are capable of matching both LSB and Dwarf Irregular and
normal high surface brightness (HSB) observed rotation
curves. Hernandez, Avila--Reese \& Firmani (2001) also find strong
evidence of a large constant density core in the dark matter halo of
the Milky Way, in comparing the results of cosmological simulations of
Milky Way formation, to extensive Galactic rotation curve
determinations. Binney \& Evans (2001)  also find that detailed studies of the Milky Way rotation curve
rule out an NFW type profile in our galaxy. 
Even in ellipticals, Romanowsky et al. (2003), through careful stellar kinematics, rule out the presence
of dark matter to the extent implied by the cuspy NFW profiles.
Since observational evidence in galaxies of all types favours
dark halos with cores, it appears reasonable to model our galactic dark halos as
King spheres. However, given the lack of a formation scenario to explain this observation, and given that the 
very successful hierarchical model implies cuspy halos, we included also one simulation using
a cuspy NFW type halo.

King halos are defined by 3 parameters: the total mass
of the halo, the total potential energy, and a shape parameter.

We take each of the dark haloes as having a velocity dispersion
$\sigma=130$~km~s$^{-1}$. A central density of
$\rho_{0}=0.02$~M$_{\odot}$~pc$^{-3}$ was assumed, in consistency with
current estimates of this quantity obtained over a range of galactic
systems (e.g. Firmani et al. 2000, Dalcanton \& Hogan 2001, Shapiro \& Ilev 2002, 
Lokas \& Mamon 2003).  We used a central potential in units of the velocity
dispersion ${\Phi_{0}/\sigma^2}=8.0$, shown by Hernandez \& Gilmore
(1998) to result in optimal rotation curves, from LSB to HSB systems.
We obtain a core radius $r_{c}=11.8$~kpc, a tidal radius $r_{tidal}=68
r_{c}$ (where the density vanishes) and a total mass $M_{halo}=1.35
\times 10^{12}$~M$_{\odot}$. The halos were modeled out to 35 core
radii, by which time 94\% of the total mass of the halo has been
included, and the dynamical times have become comparable to the Hubble
time at $z=1.5$.

For the stellar component we take a double exponential disk
given by:  

\begin{equation}
\rho(r,z)=\rho_{\star}(0)\exp(-R/R_{\star}) \exp(-|z|/z_{\star}),
\end{equation}

\noindent with the scale length, $R_{\star}$ fixed at 3.5 kpc, resulting in a
$\lambda$ parameter for the total galaxy of $\sim 0.05$. The vertical
velocity dispersion for the stars was initially adjusted to yield a
constant scale height of $z_{\star}=R_{\star}/5$. The normalization
for the stellar disk was determined using a mass Tully-Fisher relation
having a slope of 3.5, and normalized to the Milky Way
(e.g. Giovanelli et al. 1997). When changes in the mass were explored,
the disk scale lengths were scaled with the square root of the total
disk mass, (e.g. Dalcanton, Spergel \& Summers 1996; Avila-Reese,
Firmani \& Hernandez 1998), and the total disk mass scaled to the
rotation velocity in the flat regime through the above mentioned
Tully-Fisher relation.

The gas is initially distributed in an exponential disk having the
same scale height as the stars, and a scale length equal to twice the
stellar scale length, as normal spirals tend to show (e.g. Dalcanton
et al. 1996). The normalization of the gas disk is fixed by requiring
that $M_{g}=F \times M_{\star}$ i.e. that the total gas mass is $F$
times the total stellar mass. Values of 0.3 and 0.5 were explored for
this parameter, representing gas rich disks, as found by Mihos (2001b)
to be typical of disk galaxies at $z\sim1$ in a $\Omega_{\Lambda}=0.7,\Omega_{M}=0.3$ universe.  
Both the stars and gas
are given a rotation velocity in the plane of the disk to balance the
total gravitational inward pull, over which is added an isotropic 
velocity dispersion, as required by the vertical scale height 
criterion mentioned above.

GADGET uses an ideal gas equation of state, $P=\rho u (\gamma-1)$, 
where $u$ is the internal energy per unit mass, and $\rho$ is the mass 
density. One can in principle introduce an atomic cooling law, and 
trace the detailed thermal evolution of the gas component. However, 
this results in catastrophic cooling and clumping of the gas. The 
above result is natural, when one considers that the energy content of
gas in a galactic disk is heavily determined by the turbulent regime 
it is in. Once energy is found at the atomic thermal level, it is 
radiated away, for all practical purposes instantaneously. However, an 
extended ``waiting phase'' is implied by the large scale turbulence,
which is also the phase into which much of the heating processes feed 
into. Supernova explosions deposit a large fraction of their energy 
into pushing and blowing the inter stellar medium around, rather than 
representing a coherent thermal heating mechanism (e.g. Mac Low \&
Ferrara 1999; Mori, Ferrara \& Madau 2002). 

With this in mind, one can try to model the turbulence, star 
formation, supernova explosions, and feedback mechanism between the 
above, to model the gaseous component of a galaxy. This not only very 
expensive computationally, but will probably also yield a wrong
answer, given the lack of a detailed microphysics for star formation 
and turbulence. Theoretical models and direct observations of galactic
disks suggest the existence of efficient feedback regulation 
mechanisms between star formation and turbulent dissipation, capable 
of maintaining the turbulent gas at the threshold for gravitational
instability. Examples of the above can be found in Firmani, Hernandez 
\& Gallagher (1996), Martin \& Kennicutt (2001), and Silk (2001).  In
this sense, it appears reasonable to model the gas component of our 
galaxies through isothermal equations of state, assumed to be 
representative of the turbulent phase of the interstellar 
medium. Indeed, several authors have taken this approach in the 
modeling of gaseous components of galaxies (e.g. Barnes 2002;
Athanassoula \& Bureau 1999). 
 
We have thus chosen to use an isothermal equation of state,
$P=c_{s}^{2}\rho$, where $c_{s}$ is the sound speed. We incorporated 
the necessary changes to the code to make use of this, with 
$c_{s}=20$~km~s$^{-1}$ for most runs, and exploring the changes of 
this value on our final results.

We are hence assuming that heating processes, mainly shocks and
supernova explosions at all times and at all places, exactly balance
heat losses through viscous and turbulent dissipation. The above is
well justified in isolated disks, were the gas naturally oscillates
around the threshold of gravitational instability. In the more dynamic
case of interacting and merging galaxies, the assumption breaks down
whenever dynamical times become shorter than the 10 Myrs typical of
the lifetimes of massive stars. This in effect sets a time resolution
for our simulations, below which the details of our modeled mergers
are probably unreliable. However, as our main aim is to obtain the
timescale for the completion of the merger, this detail becomes of
little relevance.

\subsection{Initial conditions} \label{initial}

Each of the galaxies which are to collide is set up as indicated in
the previous subsection, with the stellar and gaseous disks introduced
inside the equilibrium dark halo. This allows the dark halo to adjust
to the presence of the disks, and it reacts by increasing the
concentration of the inner regions slightly. The above initial disk
conditions were chosen to guarantee also that the disks should be, in isolation,
stable in terms of Toomre's criterion, this was checked explicitly by 
evolving an isolated galaxy for 5 Gyr, no changes of any type were observed.

Figure~1 shows the radial profile of one of our galactic models at the
start of the simulation. The vertical axis gives the actual rotation
velocity in km/s, shown by the thin dotted curve, and the contribution
to it of the dark matter halo given by the thin dashed line. The
contribution of the exponential stellar disk is given by the solid
curve, and that of the gas by the thick dashed one. This is seen to
match inferences for the Milky Way quite well, outwards of about 3.5
kpc, (e.g. Kuijken \& Gilmore 1989; Wilkinson \& Evans 1999; Sakamoto et al. 2003). The
inclusion of a stellar bulge component, would in fact yield a much
better agreement. In any case, our modeled galaxy is seen to reproduce
observed systems quite well in having soft core dark halos, as commented
in the previous subsection.

\begin{figure}
\psfig{width=8.5cm,file=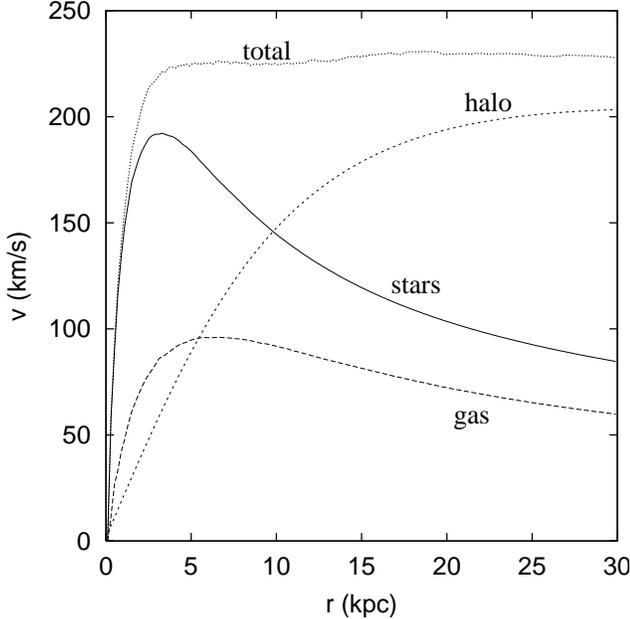,angle=0}
\caption{Rotation velocity curve, showing the total rotation curve as well as the contributions to 
this by the different components, in one of our modeled galaxies in isolation.}
\label{Evst}
\end{figure}

Once the disks are fully formed, the initial conditions for the
dynamical evolution are established, and the merger is allowed to
proceed. These initial conditions consist of a definition of the
orbital plane for the encounter, an initial separation, an impact
parameter, and the angular momentum vectors for the spins of the
disks, defined in relation to the orbital plane.

The merger is clearly defined in a parameter space having several
dimensions, the particular problem we are studying limits some of
these, the remaining we shall explore by varying the relevant
parameters.

In comparing with Press-Schechter merger trees, we have to define the
start of the merger as the moment when the two galactic halos form
part of a single bound structure. In this sense, it will be when the
centers of both galaxies are four virial radii apart.  This is based
on the result (Padmanabhan 1995) of the turnaround radius of a
fluctuation being equal to twice the virial radius, at any redshift.

Hence initially the galaxies, together with their corresponding DM
haloes, are placed at a distance $r_{i}=4 r_{virial}$, where
$r_{virial}$ is computed for a single galaxy from:

\begin{equation}
\frac{3}{4 \pi r_{virial}^{3}} \int_{0}^{r_{virial}} \rho dV=200
\rho(z_{1}),
\end{equation}
where $\rho(z_{1})=\rho_{0}(1+z_{1})^{3}$ is the critical density of
the universe at redshift $z_{1}$ (a generic redshift at which the
merger starts) and $\rho_{0}=3 \Omega_{M} H_{0}^{2}/(8 \pi G)$ is its
present value. $r_{virial}$ is always smaller than the tidal radius of the halos,
hence, the halos are fully sampled. We take $z=1.5$ and ($\Omega_{M}=0.3$,
$\Omega_{\Lambda}=0.7$ and $H_{0}=65$~km~s$^{-1}$~Mpc$^{-1}$), a
present standard set of numbers to define our cosmological scenario.

The above defines the initial separations, as a function of the
galaxies used. If one of the galaxies is taken at a different mass,
the initial separation is adjusted so that the dark halos start off
touching each other, in terms of their turnaround radii.

The orbital plane is arbitrarily taken as the XY plane, with the
galaxies staring off with zero radial relative velocities. We are
assuming they have just detached from the Hubble expansion, and only
as they begin to feel each other gravitationally will they develop an
infall radial velocity. The tangential velocity is specified through
the total $\lambda$ parameter for the full system, where:

\begin{equation}
\lambda={L \left|E \right|^{1/2} \over GM^{5/2}}
\end{equation}

Here $L$ is the total angular momentum, $E$ the total potential energy of
the system, and $M$ the total mass. Given the cosmic distribution of
$\lambda$ parameters:

\begin{equation}
P(\lambda)={1 \over {\sigma_{\lambda}(2 \pi)^{1/2}}} \exp\left[ {-\ln^{2}(\lambda/<\lambda>)}
\over{2 \sigma_{\lambda}^{2}} \right] {{d\lambda} \over {\lambda}},
\end{equation}
with $<\lambda>=0.05$ and $\sigma_{\lambda}=1.0$, (e.g.  Dalcanton et
al. (1996) and references therein), we expect the total system to have
been spun up by the surrounding tidal fields much to the same degree
as individual galaxies have. Most of the simulations were run with an
orbital $\lambda$ of 0.05, with one test at 0.025 being included. 

The remaining degrees of freedom in setting up the merger are related
to the orientation of the disks with respect to the orbital plane. We
could perhaps expect both disks to co-rotate with the orbit, as each
was spun up by a similar tidal field in the surroundings of the
forming binary system. However, the large stochastic nature of the
acquisition of angular momentum in galaxies (e.g. Catelan \& Theuns
1996) suggests that one is probably well advised not to assume any
{\it a priori} orientations. We have tested 6 different
configurations, specified by the arrows in the second column of
Table~1, where the orbital angular momentum is up.

All other initial conditions of the runs are also summarized in
Table~1.  Run C0 corresponds to a configuration where the spin of both
disks is aligned with the overall orbital motion. The orbital part of
the dimensionless parameter $\lambda$ is close to 0.05, as is that of
all other runs, except for run C$\lambda$, which is close to
0.025. The deviations from these two numbers are due to the different
orientation of the individual spins, which however make up only a
fraction of the orbital angular momentum.  In runs C0 to C5 we have
changed only the relative orientations of the two disks, with respect
to the orbital plane, with the following 12 runs being variations of
run C0 with changes in impact parameter, disk scale radii of the
galaxies, the presence and mass of a bulge, the resolution of the
simulation, the ratio of the masses of both systems, the gas fraction
of the disks, the temperature of the gas disks, total mass of each
galaxy, redshift at which the merger begins and adopted density
profile for the DM halo.

\begin{table*}
\caption{Basic parameters for each run. The table lists for each run
(labeled) the initial spin configuration, the angular momentum
parameter $\lambda$, the redshift $z$ for which the initial condition
was constructed, the total mass of the system (in units of the mass of
a single ``standard'' galaxy, used for run C0), the gas fraction with
respect to stars, the mass ratio $q$, the speed of sound $c_{s}$, the
scale radii for the stellar disks, the presence of a massive stellar 
bulge (making up 50\% and 70\% of the stellar mass in cases Cb and CbM
respectively),
and the total number of particles used in the simulation,
$N=N_{halo}+N_{star}+N_{gas}+N_{bulge}$. In all cases
$N_{halo}$=20,680, $N_{gas}=N_{star}=15,706$ and $N_{bulge}=0$, except
for run CN, where $N_{gas}=N_{star}=31,732$, and run Cb, where
$N_{bulge}=15,706$. Run H0 is analogous to run C0, but used a
Hernquist profile for the DM halo instead of a King profile (see text
for details).}
\label{parameters}
\begin{tabular}{@{}lcccccccccc} 
Run & Spins & $\lambda$ & $z$ &
$M_{t}$ & $M_{gas}/M_{stars}$ & $q$ & $c_{s}$(km~s$^{-1}$) & $R_{star}$(kpc)
& Bulge & $N$ \\ 
C0 & $\uparrow \uparrow$ & 0.082 & 1.5 & 2.0 & 0.5 &
1.0 & 20.0 & 3.50 & no & 52,092 \\ 
C1 & $\uparrow \rightarrow$ & 0.069
& 1.5 & 2.0 & 0.5 & 1.0 & 20.0 & 3.50 & no & 52,092 \\ 
C2 & $\uparrow
\downarrow$ & 0.055 & 1.5 & 2.0 & 0.5 & 1.0 & 20.0 & 3.50 & no &
52,092 \\ 
C3 & $\downarrow \downarrow$ & 0.028 & 1.5 & 2.0 &0.5 & 1.0
& 20.0 & 3.50 & no & 52,092 \\ 
C4 & $\downarrow \rightarrow$ & 0.043 &
1.5 & 2.0 & 0.5 & 1.0 & 20.0 & 3.50 & no & 52,092 \\ 
C5 & $\rightarrow
\rightarrow$ & 0.061 & 1.5 & 2.0 & 0.5 & 1.0 & 20.0 & 3.50 & no &
52,092 \\ 
C$\lambda$ & $\uparrow \uparrow$ & 0.060 & 1.5 & 2.0 & 0.5 &
1.0 & 20.0 & 3.50 & no & 52,092 \\ 
CR & $\uparrow \uparrow$ & 0.069 &
1.5 & 2.0 & 0.5 & 1.0 & 20.0 & 1.75 & no & 52,092 \\ 
Cb & $\uparrow
\uparrow$ & 0.078 & 1.5 & 2.0 & 0.5 & 1.0 & 20.0 & 3.50 & yes & 67,798
\\ 
CbM & $\uparrow
\uparrow$ & 0.078 & 1.5 & 2.0 & 0.5 & 1.0 & 20.0 & 3.50 & yes & 67,798
\\
CN & $\uparrow \uparrow$ & 0.082 & 1.5 & 2.0 & 0.5 & 1.0 & 20.0 &
3.50 & no & 84,144 \\ 
Cq & $\uparrow \uparrow$ & 0.050 & 1.5 & 1.5 &
0.5 & 0.5 & 20.0 & 3.50 & no & 52,092 \\ B & $\uparrow \uparrow$ &
0.078 & 1.5 & 2.0 & 0.3 & 1.0 & 20.0 & 3.50 & no & 52,092 \\ 
Cc & $\uparrow \uparrow$ & 0.082 & 1.5 & 2.0 & 0.5 & 1.0 & 15.0 & 3.50 & no
& 52,092 \\ 
Cm & $\uparrow \uparrow$ & 0.083 & 1.5 & 1.0 & 0.5 & 1.0 &
20.0 & 3.50 & no & 52,092 \\ 
Cmz & $\uparrow \uparrow$ & 0.075 & 2.5 &
1.0 & 0.5 & 1.0 & 20.0 & 3.50 & no & 52,092 \\ 
Cz & $\uparrow
\uparrow$ & 0.075 & 2.5 & 2.0 & 0.5 & 1.0 & 20.0 & 3.50 & no & 52,092
\\
H0 & $\uparrow
\uparrow$ & 0.075 & 1.5 & 2.0 & 0.5 & 1.0 & 20.0 & 3.50 & no & 47,670
\\
\end{tabular}

\end{table*}

\section{Results} \label{results}

\begin{figure*}
\psfig{width=15cm,file=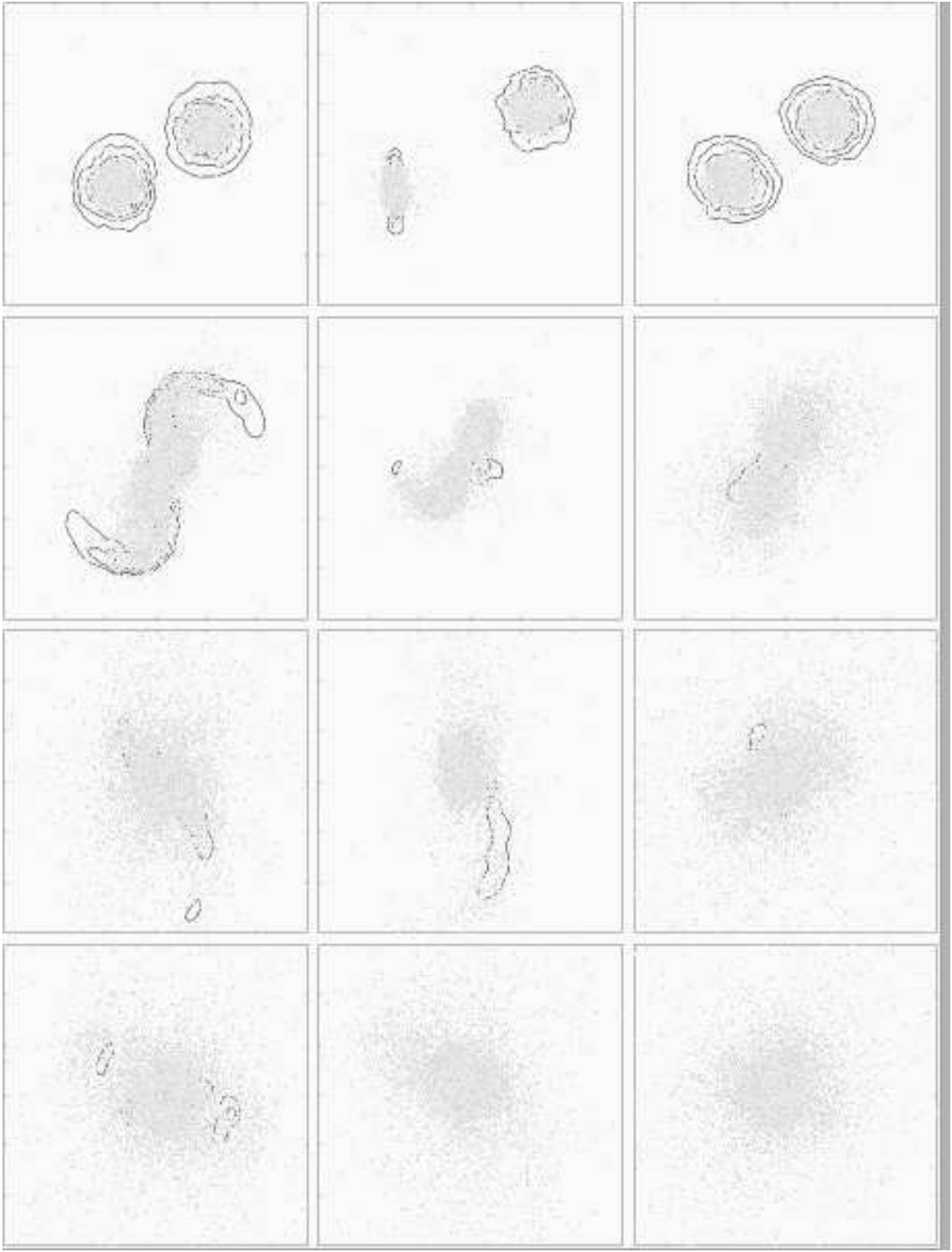,angle=0}
\caption{Snapshots of the dynamical evolution during the mergers for
runs C0 (left), C1 (center) and C3 (right), in the orbital plane, at
times $t=3.53, 4.12, 4.7, 5.3$ Gyr from top to bottom. 
The contours are logarithmic projected densities and equally spaced
every 0.5 dex, with the lowest one at
$10^{-5}$ M$_{\odot}$ pc$^{-3}$. The dots show the projected stellar
particles. Each frame is 300 kpc on a side.}
\label{rho1}
\end{figure*}

We now examine in detail a series of merger simulations,
representative of the many cases we explored. Figure~2 gives
temporal snapshots (vertical sequence) of the evolution of simulations
C0, C1 and C3, on the first, second and third columns, respectively.
The times on the horizontal rows are the same, in all cases 3.53,
4.12, 4.7 and 5.3 Gyr. The dots show the stellar component of both
disks, and the line contours give projected gas density plots, with
the much more extended dark halo particles not being shown. The
physical scale is 300 kpc on each side.

The uppermost frame in Figure~2 shows the onset of the collision, the
first 3.5 Gyr are spent by the disks in falling into each other, from
the initial condition at $z=1.5$. In case C0 both disks co-rotate with
the orbital spin, and hence are seen face on, as all frames show the
galaxies on the orbital plane. The third column also shows the
galaxies face on, as in this case both disks are counter-rotating with
the orbit, evolution up until the disks interact is hence identical to
case C0. The central column shows a simulation in which one of the
disks spins perpendicular to the orbital plane, which is seen clearly
in its first frame, where one of them is seen edge on.

\begin{figure*}
\psfig{width=15cm,file=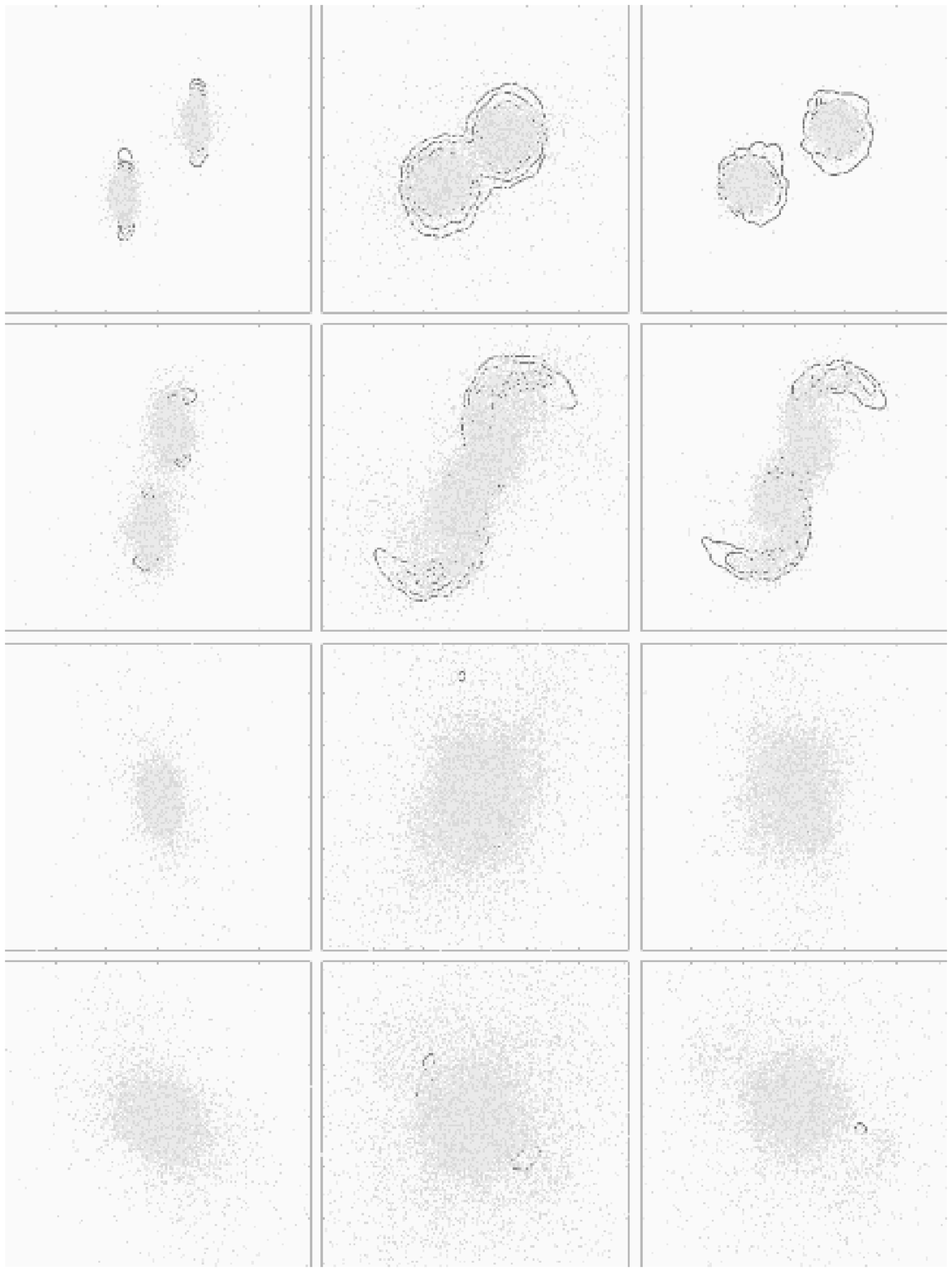,angle=0}
\caption{Snapshots of the dynamical evolution during the mergers for
runs C5, Cb and B, in the orbital plane, at times $t=3.53, 4.12,
14.7, 5.3$ Gyr from top to bottom. The
contours are logarithmic projected densities and equally spaced every 0.5 dex, with the
lowest one at $10^{-5}$~M$_{\odot}$~pc$^{-3}$. The dots show the
projected stellar particles. Each frame is 300kpc on a side.}
\label{rho2}
\end{figure*}

By the second row the first encounter has taken place, and both disks
are seen at maximum separation after the initial impact. At this stage
the morphology of the interacting systems is heavily dependent on the
initial conditions, with the large angular momentum of the disks in
simulation C0 giving rise to two well defined tidal tails, both in the
gaseous component (line contours) and in the stars (dots). The large
degree of incoherence in the angular momenta of the two disks in the
other two simulations shown leads to a significant canceling of this
component, and both gas and stars form two tight knots.

By the third row significant dissipation has taken place, especially
in the gaseous component where strong shocks and tidally induced
features develop. This now forms a dense bar in the central regions of
the remnant, no longer characterized by the double nucleus morphology
seen previously. A small tidal arm is seen in case C1, formed by
gaseous and stellar material formerly associated to the co-rotating
galaxy. By this time the differences between the three cases are
beginning to fade. The final row shows the state of the remnants at
the time when the temporal fluctuations in both total potential
energies, total kinetic energies and isophote geometry disappear, and
a final stable configuration is found. It is important to note that
the total time for this to happen is the same for the three cases
shown (see below).

Figure~3 is totally analogous to Figure~2, but shows the result of
simulations C5, Cb and B, on the first, second and third columns,
respectively. This explores the dependence of our results on yet
another different orientation, both disks perpendicular to the orbital
plane (C5).  The other two examples use the relative orientation of
case C0, but include the presence of a stellar bulge component in both
galaxies, middle column, Cb, and a significantly different gas
fraction of 0.3, in the third row, case B.

Again we see that the transition morphologies are highly sensitive to
the initial conditions, with strong tidal arms developing in any
component which co-rotates with the orbit, and falling rapidly towards
the centre when this is not the case. However, the final relaxation
times are again in excellent agreement with what was seen in the three
cases shown in Figure~2, equilibrium configurations are attained by
5.3~Gyr in all cases. Other initial orientations listed in Table~1
give analogous results.

Inspection of the frames in Figures 2 and 3 reveals that the system
has relaxed by approximately 5.2~Gyr. We proceeded to make a more
thorough estimate of this timescale by analyzing the time evolution of
various quantities for each simulation (all of them relating to the
stellar component only). The first of these involves the eigenvalues
of the inertia tensor. More specifically, we computed the square root
of the ratios of these eigenvalues ($R_1$, $R_2$, $R_3$), shown for
runs C0, C1 and C5 in Figure 4. This ratio is clearly a measure of the
relative lengths of the principal axes of the mass distribution, and
as such provides a good measurement of the overall shape of the
system. The time variation clearly shows that the strongest
fluctuations occur at the time of the first collision, with the
geometry settling down at around 5.0~Gyr.  The degree of oblateness or
prolateness is seen to depend on the initial conditions, particularly
on whether the individual galaxies co-rotate with the infall orbit. We
estimated a relaxation time from this information by
defining $\tau_{1}$ as the time by which the fluctuations in the
ratios $R_i$ (sampled at intervals of 0.2~Gyr) had dropped below 5\%.

The second quantity which we analyzed was the gravitational potential
energy. Figure 5 shows its time evolution, both for the total value
(bottom panel) and for the gas component only (top panel), for runs
C0, C1, C3, C5 and Cb, the first five of the cases shown in Figures
(2) and (3). We note that an equivalent plot for the stellar 
components is very similar (albeit at a different absolute scale) than 
what is shown for the gas value. It can be seen that the 
details of the first
maximum/minimum in the energies vary slightly from case to case, but
that in all cases no temporal oscillations remain beyond 5.3 Gyr. From
the upper panel, it is clear that the dynamical relaxation timescales
we adopt are representative of the process being modeled, and always
fall within a narrow range of values, close to 4.8 Gyr. 
The sharp dip seen in the lower
panel close to 4 Gyr shows the first collision, also seen as a sharp
dip in the upper panel. However, the strong dissipation associated
with breaking up the disks leads to a very strong loss of potential
energy (particularly in the gaseous component) as the disks merge, not
seen in the graph for the total energy, dominated by the massive dark
halos, which essentially show a rebound.

In complete analogy with the determination of $\tau_1$ described above,
we defined $\tau_2$ as the time by which fluctuations in the
gravitational potential of stars had dropped below 5\%, also sampled at
intervals of 0.2~Gyr. Taken together, these quantities estimate both a
dynamical relaxation ($\tau_{2}$), and something closely corresponding
to an ``optical relaxation'', $\tau_{1}$. The global relaxation time
was finally defined as $\tau=\max(\tau_{1}, \tau_{2})$.

The relaxation times we find depend only weakly on the
initial orientation of the disks, the details of the disk structure,
and the physical conditions within them. This suggests that our final
relaxation times are merely a dynamical result, expected to be a
function only of the free-fall times at the initial conditions. This
suggests a scaling with
$\tau\propto (1+z_{1})^{-3/2}$. We included simulation Cz to 
test this hypothesis,
a case totally analogous to C0, but set up at $z=2.5$. The results
confirmed the scaling we expected, as did cases Cm (half the total
mass at $z=1.5$) and Cmz (half the total mass at $z=2.5$). This allows
us to express the total relaxation times of cosmologically constructed
spiral galactic mergers as:

\begin{equation}
\tau_{R}={{20.6 \pm 1.86} \over {(1+z_{1})^{3/2}}} {\rm Gyr}
\end{equation}
where $z_{1}$ is the redshift at which the merger begins, and the 
range given
shows a $1 \sigma$ variation which results from the range of values we 
found in the
many experiments performed. We note that at the standard 
redshift $z=1.5$, equation (5) gives $\tau= 5.2 \pm 0.47 {\rm Gyr}$

Although no explicit star formation has been introduced, in the
interest of avoiding the introduction of free parameters pertaining to
unknown physics, we do assume implicitly a certain degree of star
formation, to justify the use of a constant gas temperature. Case Cc
(where the speed of sound is $c_{s}=15$~km~s$^{-1}$) explores the
sensitivity of our results to the actual value used for the gas
temperature, and although intermediary morphologies are slightly
affected, total relaxation times are not. Case C$\lambda$ was
calculated to explore the effect of changing the impact parameter, in
this case reduced by taking the orbital part of the $\lambda$
parameter for the system at 0.025. Again, intermediary morphologies
are affected, but final relaxation times are only slightly reduced, within the range of 
equation (5), as is the case with
changes in gas and stellar disk scale radii (case CR) and mass ratio
of the two galaxies (case Cq). Finally, case CN was a re-run of case
C0, but at double the numerical resolution for the gas and stellar
components. No differences were observed with respect to case C0,
showing that our results are robust with respect to numerical effects.
Additionally, one extra simulation, CbM tested whether including a 
very massive bulge made a substantial difference. In this case,
70\% of the mass in stars was included in the bulge, rather than 
in the more extended stellar disc. Thus in practice this simulation
contained a rather massive, compact stellar core. our determinations 
of $\tau$ were not affected beyond the values already 
established with the rest of the calculations. 

Given that cosmological N-Body simulations yield dark matter halos
having a much steeper central density profile than the king halos we
used in most of our simulations (e.g.  Navarro et al. 1996, Ghigna et al. 1998 ), 
we explore also the possible effects such a change in the dark halo might have on our
results. Case H0 is completely analogous to case C0 , but was
calculated assuming a Hernquist profile (Hernquist 1990), having the
same total mass. The scale parameter of the halo was adjusted so as to
leave the rotation curves largely unchanged. Although the central dark
matter densities in these cases were significantly larger than in the
corresponding King halo ones, as the collisions get under way, first
external and progressively more internal regions of the haloes begin
to interact and merge. This implies disturbing the orbits of dark
matter particles which make up the cusp, which is in turn dissolved
fairly quickly. What we find is that this change produces only a
marginal effect on the total relaxation times, which for case H0 fall
within the distribution of relaxation times of the King halo cases,
and indeed far from the extremes, which are due to extreme
orientations of the disks, and values of the orbital $\lambda$
parameter.
 
\begin{figure}
\hskip -210pt
\psfig{height=18cm,file=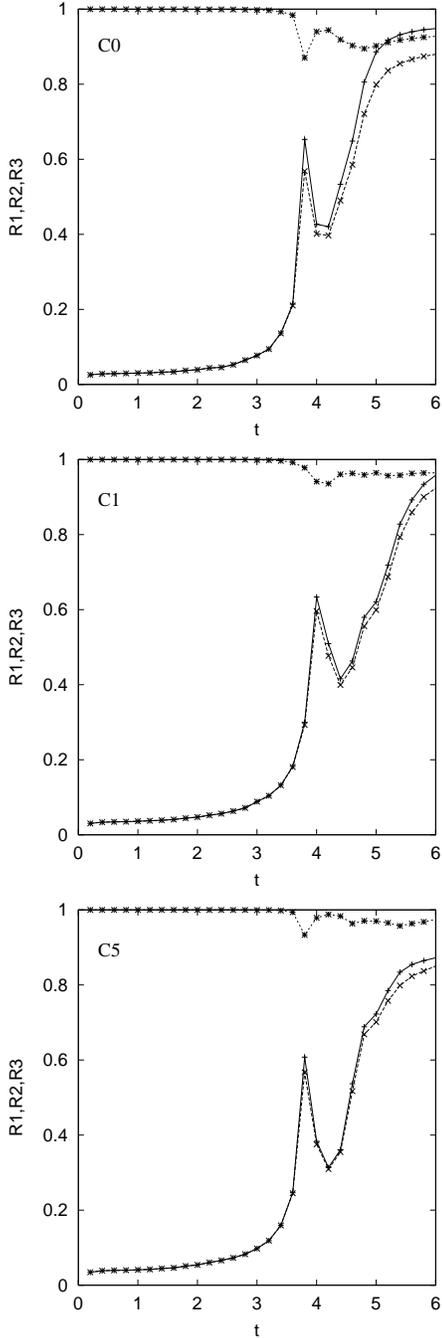,angle=0,xstretch=2.0}
\caption{Time evolution of the ratios of the principal axes of 
the inertia tensor of the stellar component for (a) run C0, (b) run C1 
and (c) run C5.}
\end{figure}

\begin{figure}
\psfig{width=8.5cm,file=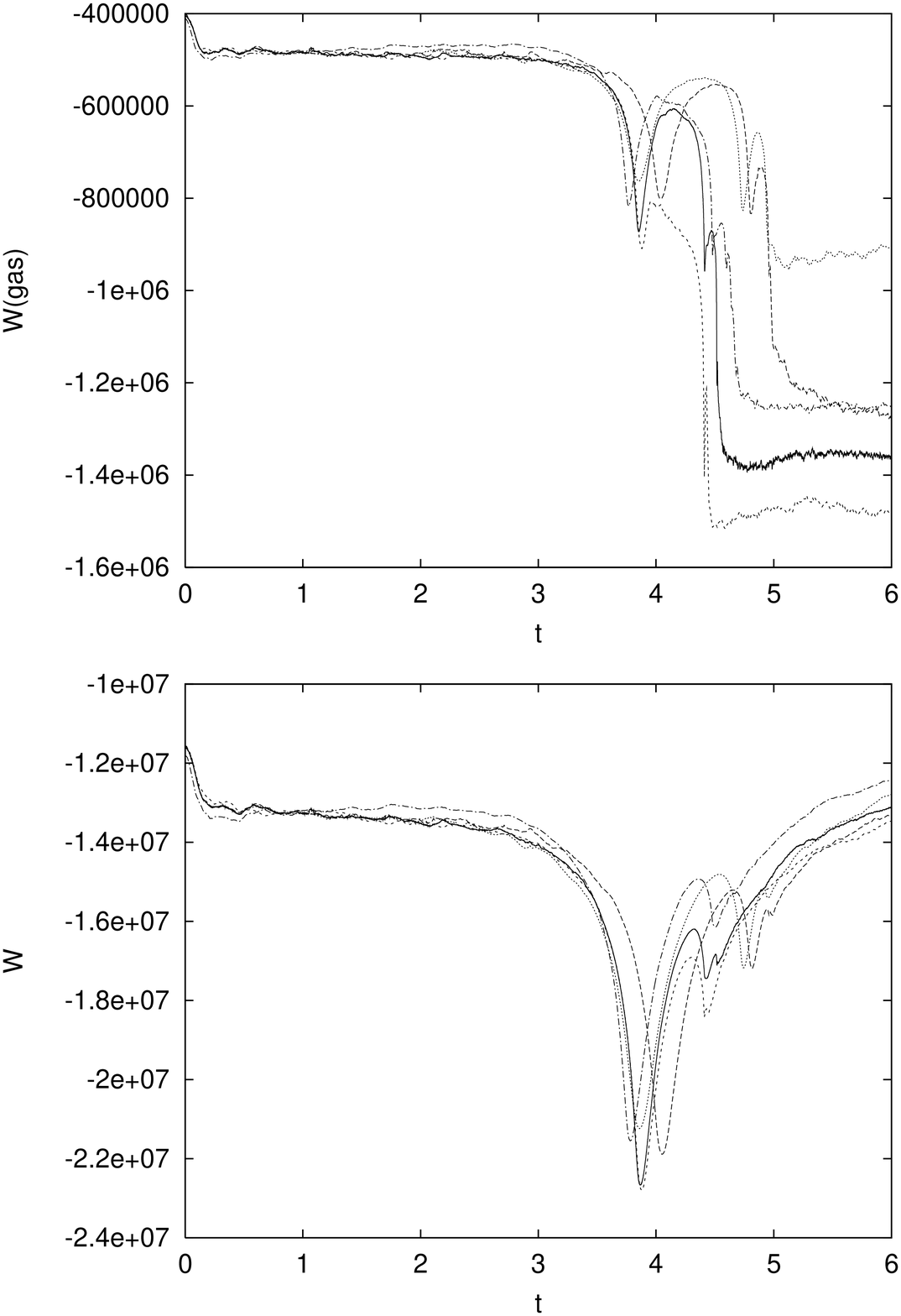,angle=0}
\caption{
Gravitational potential energies as a function of time
for runs C0, C1, C3, C5, and Cb (from top to bottom at the right end of the figure, respectively), 
gaseous components, upper panel, and
total, lower panel. Time is given in units of
$9.8 \times 10^{8}~$yr, energies in units of $2 \times
10^{53}$~erg. 
}
\label{rholong}
\end{figure}

\section{Comparison with Hierarchical merger trees}

\subsection{Analytical formulation of the problem}
In the previous section the relaxation timescale of a merger has been
established through dynamical simulations and cosmologically motivated
starting conditions, as a function of the merger redshift. We now
require an estimate of the formation timescale within the hierarchical
clustering scenario. This will be done in a rigorous fashion by
constructing merger trees within the extended Press-Schechter
formalism, including all the cosmological details. However, before
doing this it is convenient to derive an approximate analytical
solution, valid only in the simple $\Omega_{M}=1,
\Omega_{\Lambda}=0.0$ scenario, and subject to numerous simplifying
assumptions. This is done to obtain a clear understanding of the
physics of the problem, and to provide an order of magnitude estimate
to the trends and values one should expect in the more rigorous
numerical experiments.

The mass function of progenitors of an object existing at $z=z_{o}$
with mass $M_{o}$, viewed at $z=z_{1}$ will be given by:

\begin{equation}
P(M_{1})={{M_{0}} \over {(2 \pi)^{1/2}}} {{\delta_{c}(z_{1}-z_{0})}\over {(S_{1}-S_{0})^{3/2}}}
exp\left[ {{\delta_{c}^{2}(z_{1}-z_{0})^{2}} \over{2(S_{1}-S_{0})}}  \right] 
\left| { {d S_{1}}\over{d M_{1}}}  \right|
\end{equation}

In the above equation $P(M_{1})$ is the probability of finding a
progenitor of mass $M_{1}$, at $z=z_{1}$, where clearly $z_{1}>z_{0}$
(e.g. Lacey \& Cole (1993); Nusser \& Sheth (1999); Hernandez \&
Ferrara (2001)). $\sqrt{S_{i}}$ is the rms density fluctuation in a
top hat window function of radius $(3M_{i}/4\pi \rho_{0})^{1/3}$ and
$\delta_{c}$ the critical over-density for collapse, with
$\rho_{0}=3\Omega_{M} H_{0}^{2}/8 \pi G$ being the present mean mass
density of the universe.

As an example, Figure~6 shows the mass function which results from
having ``factored out'' an observed object of mass
$M=10^{12}M_{\odot}, z_{0}=1.9$, amongst its progenitors at various
higher redshifts, $z=z_{1}$. This figure was calculated within a full
$\Lambda CDM$ scenario, the results for a simplified cosmology being
qualitatively identical. We see that as $z_{1}$ approaches $z_{0}$,
the mass function of progenitors tends to a delta function at
$M_{1}=M_{0}$, as was to be expected. At progressively higher
redshifts, the mass functions are characterized by very well defined
maxima, which shift to progressively lower masses (note the
logarithmic scales on both axes). In this way, we see that for the
above case, in going to redshifts $ \geq 3.9$ (dashed curve), the chance of
finding a progenitor with a mass comparable to, say, $M_{0}/3$,
sharply drops.

For the analytical calculation, we shall take a fixed $\delta_{c}=1.7$
and $\Omega_{M}=1.0$, as well as:

\begin{equation}
S(M_{i})=S_{8} M_{i}^{-1/3},
\end{equation}
valid for an effective spectral index in the galactic region of
$n=-2$, where $S_{8}$ is a normalizing factor to be fixed later
(e.g. Padmanabhan 1995).

The idea now is to identify the redshift $z_{1Max}$ which corresponds
to $dP(M_{1})/dM_{1}=0$ evaluated at $M_{1}=M_{0}/2$, as the redshift
interval previous to $z_{0}$, during which a major merger might have
formed the object ($M_{0},z_{0}$), as further back in the past of this
redshift, the chances of finding a progenitor having half the mass of
the observed object drop abruptly.

Substituting the power law dependence for $S(M_{i})$ in the rather
cumbersome expression for $dP(M_{1})/dM_{1}=0$, evaluated at
$M_{1}=M_{0}/2$ yields:

\begin{equation}
(z_{1}-z_{0})={1 \over \delta_{c}} \left( {{S_{8}}\over {M_{0}^{1/3}}}
\right)^{1/2}
\label{deltaz}
\end{equation}

We can now evaluate $S_{8}$ from $S_{8} M_{8}^{-1/3} =1$, with:

\begin{equation}
M_{8}= {{4 \pi} \over {3}} (8 \times 10^{3} {\rm kpc})^{3} \rho_{c},
\end{equation}
yielding:

\begin{equation}
S_{8}=8\times 10^{4} \Omega_{M}^{1/3}.
\end{equation}
i.e., a $\sigma_{8}=1$ normalization for the spectrum.

Substituting this last result, and the numerical value 
for $\delta_{c}$ into equation~(\ref{deltaz}) yields, 

\begin{equation}
\Delta z_{M}=166 \left( {\Omega_{M} \over M_{0}}   \right)^{1/6}
\end{equation}

The above expression gives the redshift interval, backwards of a
redshift of observation, $z_{0}$, beyond which it is unlikely that a
major merger could have occurred, resulting in the observed object of
mass $M_{0}$, and defines the merger timescale.  It is interesting
that in the simplified cosmology taken for this calculation, no
explicit dependence on $z_{0}$ remains.

Given that the initial conditions for the merger occurring at
$z=z_{1}$ require that the two galaxies be placed at a separation of a
multiple of their current virial radii, the relaxation timescale,
$\tau_{R}$ will be estimated here as a multiple $\alpha$ of order
unity of the free fall timescale for a system having 200 times the
background density of the universe. hence,

\begin{equation}
\tau_{R}=\left( {{3 \pi \alpha} \over {32}}\right) ^{1/2}
\left( {{1} \over {200 \rho_{0} G}}\right)^{1/2} (1+z_{1}) ^{-3/2}
\end{equation}

Introducing numerical values, and using $t_{H}=(2/3) H_{0}^{-1}$ for
the present age of the universe gives:

\begin{equation}
\tau_{R}= 0.17 \alpha {{t_{H}} \over {(1+z_{1})^{3/2} \Omega_{M}^{1/2}}}
\end{equation}

Before comparing the above relation to the merger timescales of
equation(11), we shall use:

\begin{equation}
1+z_{i}=t_{H}^{2/3} t_{i}^{-2/3}
\end{equation}

to arrive at:

\begin{equation}
166 \left( {\Omega_{M} \over M_{Lim}(z_{0})}   \right)^{1/6}=
(1+z_{0}) \left[ \left(1+ {{0.17 \alpha} \over {\Omega_{M}^{1/2}}} \right)^{2/3} -1 \right],
\end{equation}
for the mass $M_{Lim}(z_{0})$ which at $z=z_{0}$ has a merger
timescale equal to the relaxation timescale at the point at which the
merger began. Introducing $\Omega_{M}=0.3$ and $\alpha=1.6$ yields:

\begin{equation}
M_{Lim}(z_{0})=6.4 \times 10^{15} (1+z_{0})^{-6} M_{\odot}.
\end{equation}

The above equation implies that at each redshift $z_{0}$, there should
be a maximum limit mass above which an observed system could not
possibly look like a relaxed elliptical galaxy, as then the relaxation
timescale becomes longer than the merger timescale, and the object
should look like an interacting system. One can note that this holds
even at $z=0$, however, at a fairly large total mass of $6.4 \times
10^{15}M_{\odot}$, which should be divided by a factor of about 20 to obtain
stellar masses. Still, this limit mass is seen to fall very rapidly as
$(1+z_{0})^{-6}$, and for example, for $M=2 \times 10^{12}$ ---in the
range of estimates for the mass of our Galaxy (e.g. Wilkinson \& Evans
1999)--- we arrive at $z_{0}=2.8$ as the limit
beyond which elliptical galaxies of that mass cannot be explained as
having originated in a major merger. The choice of $\alpha$ was
motivated by the results of the simulations performed here, and that
of $\Omega_{M}$ by an attempt to come closer to a more realistic
$\Lambda CDM$ scenario. This last choice introduces a degree on
inconsistency in the above calculation, which is only valid for
$\Omega_{M}=1$. However, as this is only intended as a conceptual
guide, the result is probably a good first approximation.

\subsection{Full $\Lambda CDM$ comparisons}

\begin{figure}
\psfig{width=8.5cm,file=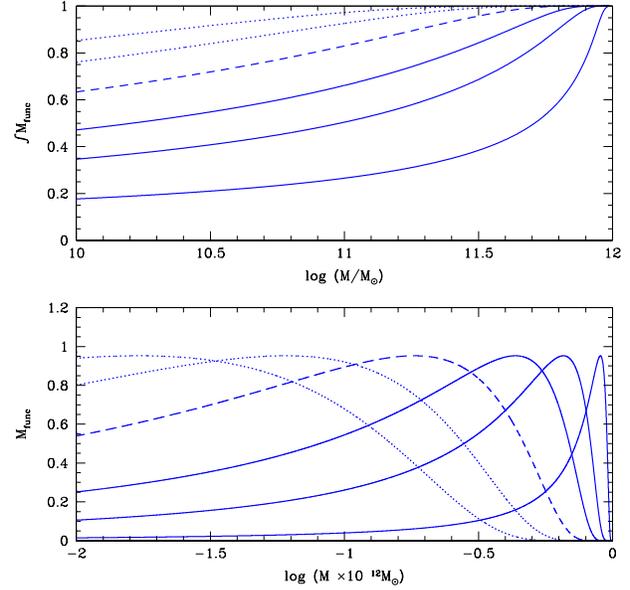,angle=0}
\caption{{\bf Lower panel:} Mass functions of progenitors of a $10^{12} M_{\odot}$ system,
observed at $z=1.9$, at different previous redshifts: 2.4, 2.9 and 3.3, (solid curves), 3.9,
(dashed curve), and 4.5 and 5.1 (dotted curves). All curves have been normalized to 1.
{\bf Upper panel:} Integral of the mass functions shown in the lower panel. It can be seen that for 
redshifts higher than 3.9 (dotted curves), the chances of obtaining one progenitor with a mass
of $0.3 \times 10^{12} M_{\odot}$ or larger, drop below 5\%, with the probability of two such
fragments occurring being well below 0.25\%.}
\label{rho2}
\end{figure}

We shall now turn to more precise numerical calculations, the results
of which can be understood more clearly by using the above results as
a conceptual guide, and an order of magnitude estimate.

The lower panel of Figure~6 gives the mass functions of progenitors of
an object of total mass $10^{12} M_{\odot}$, observed at $z_{0}=1.9$,
at various higher redshifts $z_{1}=$2.4, 2.9 and 3.3 (solid curves),
3.9 (dashed curve), and 4.5 and 5.1 (dotted curves). This time the
calculation was performed numerically evaluating equation (6) using a
full $\Lambda CDM$ scenario, with a fluctuation spectrum taken from
Percival \& Miller (1999). These curves are characterized by a maximum
located at a value of the fragment mass which is a monotonically
decreasing function of redshift. In this way, if the redshift at which
the mass function of progenitors is calculated is close to the
redshift of observation, the mass function becomes increasingly
dominated by an object having a mass very close to that which the object
has at $z=z_{0}$.

At redshifts much higher than the observation redshift, the mass
function becomes dominated by progenitors having only a few hundredths
of the mass of the observed object, e.g. the dotted curves for
$z_{1}=4.5, 5.1$. It is clear that if we want to form the $10^{12}
M_{\odot}$, observed at $z_{0}=1.9$ from a major merger, this cannot
have taken place at such high redshifts. The solid curves show that
this hypothetical major merger could well have taken place at a
redshift of 2.4, 2.9 or 3.3, as at these values the mass function of
progenitors for our $10^{12} M_{\odot}$, $z_{0}=1.9$ object includes a
high probability of finding objects in the $(0.3-1.0) \times 10^{12}
M_{\odot}$ range. This is clearly seen in the top panel, where curves
corresponding to those of the bottom one give the integrals of the
mass functions, normalized to 1.0, and hence the functions to be
sampled if the progenitors of our test object are to be obtained, at
any of the previous redshifts shown.

Although the mass functions at $z_{1}$ of 2.4, 2.9 or 3.3 do imply a
high probability of finding progenitors in the ``major merger'' range,
we cannot say that if the object $10^{12} M_{\odot}$, $z_{0}=1.9$ is
an elliptical galaxy it was formed by a major merger having occurred
in this redshift range. This is because in this redshift range,
equation (5) implies that the relaxation timescale for such a merger,
$\tau_{R}(z_{1})$, is larger than the time intervals between any of
those redshifts and $z_{0}=1.9$. This shows that if a major merger
occurred between $z_{1}=1.9$ and $z_{1}=3.3$, the result at
$z_{0}=1.9$ would be an interacting system, and not a relaxed
elliptical galaxy. In fact, for this system, only just at the redshift
$z_{1}=3.9$, corresponding to the dashed curve, does the relaxation
timescale become equal to the time interval $z_{0}-z_{1}$. This
redshift $z_{1}=3.9$ also corresponds to the maximum redshift out to
which there is some chance of obtaining a progenitor in the major
merger range, defined as at least a 5\% chance of obtaining one
progenitor with a mass upwards of 0.3 times the mass at the
observation redshift. This condition defines the merger timescale,
$\Delta t_{M}$, which in this case is equal to $\tau_{R}$.

\begin{figure}
\psfig{width=8.5cm,file=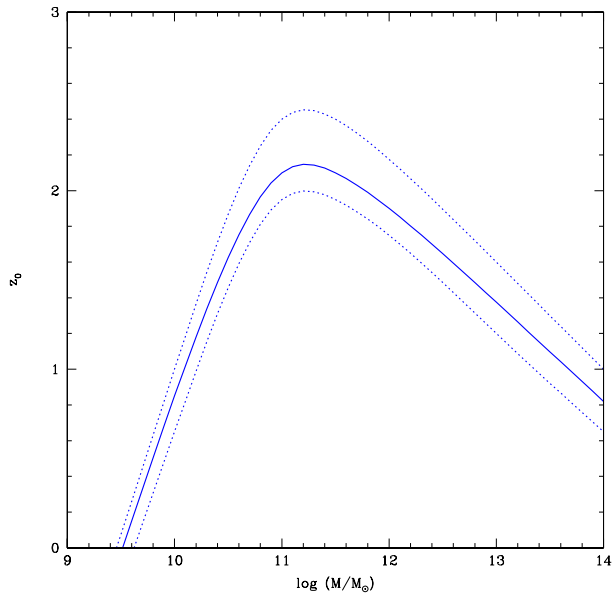,angle=0}
\caption{$z_{0lim}(M)$, the maximum observation redshift at which an elliptical galaxy of 
total mass $M$ can be thought of as having been formed as the result of a recent major merger.}
\label{rho2}
\end{figure}

In this way, we identify the observation redshift of 
$z_{0}=1.9$ as the maximum redshift at which an elliptical galaxy of
mass $10^{12} M_{\odot}$ can be observed, if it is to be thought of as
having been formed by a major merger. Smaller observation redshifts at this mass
result in cases where $\tau_{R} < \Delta t_{M}$, and hence suitable
candidates for a major merger origin, if they are elliptical
galaxies. On the other hand, larger observation redshifts at this mass satisfy the
condition $\tau_{R} > \Delta t_{M}$, and would therefore look like
interacting systems, if formed by a major merger.

We can now repeat the calculation, and look for the limit observation
redshift, $z_{0lim}(M)$ above which the major merger scenario fails for
elliptical galaxies, as a function of total mass.  This was done in
constructing Figure~7, which shows $z_{0lim}(M)$, as a function of
mass, solid curve. The dotted curves represent our intrinsic
uncertainty range for this quantity, given the variety of relaxation
timescales we obtained, for the large range of orientations, dark
matter profiles, orbital $\lambda s$ and disk physics considered.
Benson et al. (2002) reach a simplified version of our criteria, in requiring more than 10
relaxation timescales to have elapsed between the formation and observation of 
high redshift elliptical systems, for colour and colour gradients data to be consistent with 
the hierarchical scenario.

\section{Conclusions and Discussion}

In considering our results of Figure~7 we firstly note that the
expected scaling obtained for the simple analytic case is found only
at high masses, where the curve of $z_{0lim}(M)$ does indeed tend to a
$M^{-1/6}$ scaling. However, the slight curvature found in the more
realistic power spectrum, together with the high structure formation
redshift in the $\Lambda CDM$ scenario, result in a downturn at lower
masses for this curve. This is interesting, as it establishes a
maximum redshift of $z_{0}=2.5$ above which any observed elliptical
galaxy, whatever its mass, falls above the $z_{0}=z_{0lim}(M)$ curve,
i.e. in the $\tau_{R} > \Delta t_{M}$ region.  Any such elliptical
galaxy would look like an interacting system, if it had formed as the
result of a major merger.

Secondly we note that the region below the curve, where $\tau_{R} <
\Delta t_{M}$, encompasses the majority of observed elliptical
galaxies, which can hence be thought of as having been formed by the
merger mechanism. However, the downturn at lower masses identifies a
local maximum mass of $1.3 \times 10^{10} M_{\odot}$ ($6.3 \times
10^{8} M_{\odot}$ in baryons), below which any observed ellipticals at
any observation redshift $z_{0}>1.0$, must be though of as having
arisen through some mechanism distinct to the mayor merger hypothesis. The
corresponding limit at $z_{0}=0$ becomes $3 \times 10^{9} M_{\odot}$
($1.6 \times 10^{8} M_{\odot}$ in baryons), comparable to the baryonic
mass of many local dwarf ellipticals and larger than the few $ \times
10^{7} M_{\odot}$ in baryons inferred for the dwarf spheroidal
satellites of the Milky Way.

Again, we note the existence of a maximum redshift of
$z=2.15 \pm0.25$ beyond which all galaxies fall in the $\tau_{R} >
\Delta t_{M}$ region.A number of recent studies have shown the
existence of observed ellipticals in this region, inferred to be
relaxed normal ellipticals though multi-wavelength studies of their
stellar populations, or the normal appearance of their
spectra. Studies of colours, luminosity function evolution, evolution (or lack thereof)
of the fundamental plane zero point and clustering properties of high redshift
ellipticals ($1<z<3$) have shown that little evolution is seen to
have occurred through merging over that period in field elipticals, and that, at those high
redshifts, most systems ($\sim 70\%$) are allready relaxed normal ellipticals,
even considering problems related to dust obscuration, often at odds with the details of what the 
standard hierarchical
scenario predicts e.g. Im et al. (1996), Moriondo et al. (2000), 
Daddi et al. (2000), Brinchmann \& Ellis (2000), Daddi et al. (2001), Daddi et al (2002), 
Cimatti et al (2002), Im et al. (2002), and Miyazaki et al. (2002).

This last point forms our strongest conclusion. There is direct
observational evidence for the existence of elliptical galaxies for
which the condition $\tau_{R} > \Delta t_{M}$ is met, and which within
the hierarchical scenario of structure formation, in the
observationally constrained $\Lambda CDM$ scenario, (within the assumption
of ellipticals forming in major mergers of spirals) should look like
interacting systems, and not like relaxed elliptical galaxies at all.

It must
be noted that we have used essentially present day galaxies and
present day galactic scalings to model each of the spirals in the
merger simulations. This is not entirely consistent, as our merger
simulations take place at redshifts of 1.5 and 2.5. There is
considerable evidence, both theoretical, within the framework of
hierarchical structure formation models (e.g. Avila-Reese et al. 1998)
and observational (e.g. Vogt 2001; Vogt \& Phillips 2002) which
implies very little or indeed no evolution of the mass Tully-Fisher
relation with redshift. In this respect, the use of local values for
this important structural relation for our high redshift galaxies is
well justified.

The use of local disk scale length vs. disk mass relation however, is
a different matter. Again both theoretical and observational studies
of the redshift evolution of this scaling agree, giving a strong
reduction in the disk scale length, in going to higher redshifts. This
last point was not considered in most of our the simulations, because including it
would only lead to longer relaxation timescales, $\tau_{R}$, and hence
more dramatic, lower values of $z_{0lim}(M)$, at all masses. This is
clear if one considers that in reducing the typical sizes of disks,
one is limiting the action and effects of the tidal forces which bring
about dissipation and relaxation in the merging galaxies, as these
dynamical effects scale with the size of the objects upon which they
act. In a limiting case, one can imagine very small and compact disks
which could be treated as point masses, the ``merger'' would not be
more than the formation of a binary system in mutual orbit. Indeed,
this effect starts to appear once the typical disk sizes become
smaller than the distances of closest approach, relaxation times get
longer and eventually tend to infinity, as the components become
smaller. An analogous way of viewing this effect is to consider the
typical density of the components. From a simplified classical tidal
criterion, one can expect components to survive if their
characteristic densities are higher than the average densities within
their orbits. Indeed, our runs with shorter disk scale lengths yielded
somewhat larger relaxation timescales.

It is hence clear that the use of local scaling laws leads to a
conservative estimate of $z_{0lim}(M)$, with results for a fully self
consistent hierarchical scenario of structure formation yielding much
more restrictive and lower values of $z_{0lim}(M)$, at all masses. The
same can be said of the lack of a specific recipe for star formation,
the introduction of which would result in the conversion of gas into
stars i.e. of a dissipative component into a non-collisional
ingredient, hence lengthening the relaxation times. This last effect
would be enhanced in the case of a starburst regime triggered by the
pile up of gas in the central regions, and the subsequent loss of gas
trough a galactic wind, also increasing the fraction of the
non-collisional stellar component.

Finally, we note that the arguments presented here are not the only
objections that have been raised against the idea of all ellipticals
being the result of major mergers. An early example mentioning many of the
principal objections can be found in Ostriker (1980), who noted that the tight colour-magnitude
and metallicity-magnitude relations seen in ellipticals would be hard to justify in a scenario
dominated by the random assemblage of smaller spirals. He also pointed to the deeper potential wells found in
ellipticals than in spirals, and the dissipationless nature of gravitational processes as a difficulty
to this scenario, which also would have a hard time explaining the smaller physical scale lengths
of ellipticals over spirals, since dissipationless mergers would tend to increase the scale of a system.
Many of these objections have been stated again since in more careful terms, some examples follow.

Wyse (1998) has pointed out that
the high phase-space density seen in elliptical galaxies and early
bulges is incompatible with the formation of these systems out of the
dissipational merger of stellar disks. This last point can be
alleviated by the introduction of a dissipational component, such as
the gaseous disk, coupled gravitationally to the stellar component.
However, the very red colours of elliptical galaxies at high redshift
limits the amount of gas that can be included in such a merger, the
details depending on the efficiency with which a galactic wind could
subsequently clear the merger of gas. Mihos (2001) finds that
ellipticals which fall into anomalous places in the central parameter
relationship invariably appear as merger remnants, and concludes that
if mergers are the formation mechanisms for elliptical galaxies, these
must have taken place at very high redshift. Interacting and starburst
systems seen at moderate redshifts cannot form a significant fraction
of the local elliptical population.

Mihos \& Hernquist (1996) raise the problem of what happens to the gas in merging spirals,
and point out that the extreme infall towards the centre seen in some simulations, if accompanied
by in situ star formation, would result in light profiles for ellipticals which would be 
too centrally cusped to accommodate observed de Vaucouleur's type profiles.
Ellis et al. (2001b) find that the relative colours of bulges and ellipticals in the redshift
range $0.6<z<1.0$ are at odds with the predictions of the hierarchical scenario, suggesting
very early formation epochs for ellipticals, or substantial ``rejuvenation'' of spiral bulges.
Pozzetti et al. (2003) in fact, find that a careful study of the evolution of the K-band
luminosity function agrees more with simple passive evolution models than with cosmologically
motivated hierarchical clustering scenario of Kauffmann (1999). Firth et al. (2002) conclude
that the Las Campanas redshift survey rules out models where only passive evolution plays a part, but
also that data are inconsistent with the low formation epochs of field ellipticals predicted in
hierarchical scenarios.
Birchmann \& Ellis (2000) also note that the space densities of both large ellipticals and
large spirals change very little out to $z\sim 1$, making it unlikely that large spirals
are merging to form large ellipticals.

In going to spiral galaxies, studies of the velocity dispersion tensor in the Milky Way and other
late type spirals e.g. Binney (2001) show that the lack of strong discontinuities
in the vertical velocity of stars as a function of age appear to contradict the merger
scenario, with dynamical heating by spiral arms being sufficient to account for the observations. Labbe
et al. (2003) obtained deep imaging in the IR of HDFS objects, and found that 50\% of the brightest objects, 
which in the optical show up as knotty mergers, are actually normal, relaxed large disks at $1.4<z<3.0$

There seems to be mounting evidence pointing to a high formation
redshift for ellipticals (and possibly also spirals), the results of our present study offering
new support and being in accordance with the conclusions of a variety
of independent studies centering on widely different aspects of the physics of the problem. 
If large ellipticals did not form out of the merger of comparably sized spirals, might they have formed out of
the merger of comparably sized ellipticals? The relaxation timescales would be much longer than 
for the spiral mergers, as ellipticals lack the dissipative gaseous component, and have smaller
sizes which make tides less relevant, making the problem we have pointed out in this paper worse.
As alternatives we might think of many small spirals/ellipticals being swallowed by a growing 
elliptical system over a few Gyr period, provided most of the action ended before $z \sim 2$.
It is perhaps time to consider what
modifications are needed in the present galactic assemblage scenario
in order to make it compatible with a high formation redshift for
elliptical galaxies.

\section*{Acknowledgments}

The authors wish to thank T. Padmanabhan and C. Firmani for helpful
discussion, the Institute of Astronomy, Cambridge for its
hospitality during the final phases of this work and the referee, Fabio Governato, for 
a detailed revision and helpful corrections which improved the final version.  Support for this
work was provided by CONACyT (27987E) and (I39181-E) and DGAPA--UNAM
(IN-110600).

\label{lastpage}


\begin{thebibliography}{}

\bibitem{} Athanassoula E., Bureau M., 1999, ApJ, 522, 699

\bibitem{} Avila-Reese V., Firmani C., Hernandez X., 1998, ApJ, 505, 37

\bibitem{} Barnes J. E., 2002, MNRAS, 333, 481

\bibitem{} Baugh, C. M., Cole, S., Frenk, C. S., 1996, MNRAS, 283, 1361 
  
\bibitem{} Bendo G. J., Barnes J. E., 2000, MNRAS, 316, 315

\bibitem{} Benson, A. J. Ellis, R. S., Menanteau, F., 2002, MNRAS, 336, 564

\bibitem{} Binney J. J., 2001, ASP Conf. Ser. 230, 63

\bibitem{} Binney, J. J., Evans, N. W., 2001, MNRAS, 327, 27

\bibitem{} Brinchmann J., Ellis R. S., 2000, ApJ, 536, L77

\bibitem{} de Blok W. J. G., McGaugh S. S., 1997, MNRAS, 290, 533

\bibitem{} de Blok W. J. G., McGaugh S. S., Rubin V. C., 2001, AJ, 122, 2396

\bibitem{} de Blok W. J. G., Bosma A., McGaugh S., 2003, MNRAS, 340, 657

\bibitem{} Burkert A., Silk J., 1997, ApJ, 488, 55

\bibitem{} Burkert A., Naab T., 2003, astro-ph/0305076

\bibitem{} Catelan P., Theuns T., 1996, MNRAS, 282, 455

\bibitem{} Cimatti A., et al., 2002, A\&A, 391, L1

\bibitem{} Daddi E., Cimatti A., Renzini A., 2000, A\&A, 362, L45

\bibitem{} Daddi E., Broadhurst T., Zamorani G., Cimatti A., Röttgering H., Renzini A.,
2001, A\&A, 376, 825

\bibitem{} Daddi E. et al., 2002, A\&A, 384, L1

\bibitem{} Dalcanton J. J., Hogan C. J., 2001, ApJ, 561, 35
 
\bibitem{} Dalcanton J. J., Spergel D. N., Summers F. J., 1996, ApJ, 482, 659

\bibitem{} Ellis R., 2001, astro-ph/0102056

\bibitem{} Ellis R. S., Abraham R. G., Dickinson M., 2001b, ApJ, 551, 111

\bibitem{} Firmani C., Hernandez X., Gallagher J., 1996, A\&A, 308, 403

\bibitem{} Firmani C., D'Onghia E. D.,  Avila-Reese V., Chincarini G., Hernandez X., 
2000, MNRAS, 315, L29

\bibitem{} Firmani C., D'Onghia E. D., Chincarini G., Hernandez X., Avila-Reese V.,
2001, MNRAS, 321, 713

\bibitem{} Ghigna S., Moore B., Governato F., Lake G., Quinn T., Stadel J., 1998, MNRAS, 300, 146
   
\bibitem{} Ghigna S., Moore B., Governato F., Lake G., Quinn T., Stadel J., 2000,
ApJ, 544, 616

\bibitem{} Gingold R. A., Monaghan J. J., 1977, MNRAS, 181, 375

\bibitem{} Giovanelli R., Haynes M. P., Herter T., Vogt N. P. 1997, AJ, 113, 22 

\bibitem{} Gnedin O. Y., Zhao H., 2002, MNRAS, 333, 299

\bibitem{} Hernandez X., Avila-Reese V., Firmani C., 2001, MNRAS, 327, 329

\bibitem{} Hernandez X., Ferrara A., 2001, MNRAS, 324, 484

\bibitem{} Hernandez X., Gilmore G., 1998, MNRAS, 294, 595

\bibitem{} Hernquist L., 1990, ApJ, 356, 359

\bibitem{} Hibbard J., Mihos C., 1995, AJ, 110, 140

\bibitem{} Im M., Griffiths R. E., Ratnatunga K. U., Sarajedini V. L.,
1996, ApJ, 461, L79

\bibitem{} Im et al., 2002, ApJ, 571, 136

\bibitem{} Kauffmann, G., 1996, MNRAS, 281, 487

\bibitem{} Kauffmann G., Colberg J. M., Diaferio A., White S. D. M., 1999, MNRAS, 303, 188

\bibitem{} Khochfar, S., Burkert, A., 2003, Ap\&SS, 285, 211
 
\bibitem{} King I. R., 1966, AJ, 71, 64

\bibitem{} Kuijken K., Gilmore G., 1989, MNRAS, 239, 571

\bibitem{} Labbe I., et al., 2003, ApJ, 591, L95

\bibitem{} Lacey C., Cole S., 1993, MNRAS, 262, 627

\bibitem{} Lokas, E. L., 2002, MNRAS, 333, 697
    
\bibitem{} Lokas, E. L.,  Mamon, G. A., 2003, MNRAS, 343, 401
   
\bibitem{} Lucy L. B., 1977, AJ, 82, 1013

\bibitem{} Mac Low M., Ferrara A., 1999, ApJ, 513, 142

\bibitem{} van der Marel R. P., Magorrian J., Carlberg R. G., Yee H. K. C., Ellingson E.,
2000, AJ, 119, 2038

\bibitem{} Martin C. L., Kennicutt R. C., 2001, ApJ, 555, 301

\bibitem{} McGaugh, S. S., Rubin, V. C., de Blok, W. J. G. 2001, AJ, 122, 2381

\bibitem{} Mihos J. C., Hernquist L., 1996, ApJ, 464, 641

\bibitem{} Mihos C., 2001, ASP Conf. Ser. 240, 143

\bibitem{} Mihos C., 2001b, ASP Conf. Ser. 230, 491

\bibitem{} Miyazaki et al., 2002, astro-ph/0210509 

\bibitem{} Monaghan J. J., 1992, ARA\&A, 30, 543

\bibitem{} Moore et al., 1999, ApJ, 524, 19

\bibitem{} Mori M., Ferrara A., Madau P., 2002, ApJ, 571, 40

\bibitem{} Moriondo G., Cimatti A., Daddi E., 2000, A\&A, 364, 26

\bibitem{} Navarro J., Frenk C., White S., 1996, ApJ, 462, 563

\bibitem{} Navarro J., Steinmetz M., 2000, ApJ, 538, 477

\bibitem{} Nusser A., Sheth R. K., 1999, MNRAS, 303, 179

\bibitem{} Ostriker, J P., 1980, ComAp, 8, 177

\bibitem{} Padmanabhan T., 1995, Structure Formation in the Universe, 
Cambridge University Press, Cambridge

\bibitem{} Percival W., Miller L., 1999, MNRAS, 303, 179

\bibitem{} Pozzetti, L., et al., 2003, A\&A, 402, 837

\bibitem{} Romanowsky A. J., Douglas N. G., Arnaboldi M., Kuijken K., Merrifield M. R., 
Napolitano N. R., Capaccioli M., Freeman K. C., 2003, astro-ph/0308518
  
\bibitem{} Sakamoto T., Chiba M., Beers T. C., 2003, A\&A, 397, 899

\bibitem{} Shapiro P. R., Iliev I. T., 2002, ApJ, 565, L1 

\bibitem{} Silk J., 2001, MNRAS, 324, 313

\bibitem[\protect\citename{Springel, Yoshida \& White }2001]{syw01}
   Springel, V., Yoshida, N., White, S. D. M., 2001, New Ast., 6, 79

\bibitem{} Somerville R. S., Primack J. R., Faber S. M., 2001, MNRAS, 320, 504

\bibitem{} Steinmetz, M., Navarro, J. F., 2002, NewA, 7, 155

\bibitem{} Tyson J. A., Kochanski G. P., dell'Antonio I. P., 1998, ApJ, 498, L107

\bibitem{} Vogt N. P., 2001, Deep Fields, ESO/ECF/STScI workshop,
p.112

\bibitem{} Vogt N. P., Phillips A. C., 2002, AAS, 200 4008

\bibitem{} White S. D. M., Frenk C., Davis M., Efstathiou G., 1987, ApJ, 313, 505

\bibitem{} White S. D. M., Rees M. J., 1978, MNRAS, 183, 341
 
\bibitem{} Wilkinson M. I., Evans W., 1999, MNRAS, 310, 645

\bibitem{} Wyse R. F. G., 1998, MNRAS, 293, 429


\end{thebibliography}
\end{document}